\documentclass[amsmath, aps, pra, twocolumn, floatfix]{revtex4-2}
\usepackage{graphicx}

\begin{document} 
	
\title{Optical force between two coupled identical parallel optical nanofibers}
	
\author{Fam Le Kien}
\affiliation{Okinawa Institute of Science and Technology Graduate University, Onna, Okinawa 904-0495, Japan}
		
\author{S\'{i}le Nic Chormaic}
\affiliation{Okinawa Institute of Science and Technology Graduate University, Onna, Okinawa 904-0495, Japan}
	
\author{Thomas Busch}
\affiliation{Okinawa Institute of Science and Technology Graduate University, Onna, Okinawa 904-0495, Japan}
	
\date{\today}
	
\begin{abstract}		
We study the optical force between two coupled parallel identical nanofibers using the rigorous array mode theory. We show that the forces of the even array modes are attractive, while the forces of the odd array modes are repulsive. We examine the dependencies of the optical forces on the array mode type, the fiber radius, the light wavelength, and the fiber separation distance. We show that, for a given power and a given separation distance, the absolute value of the force achieves a peak when the fiber radius and the light wavelength are appropriate.
\end{abstract}
	
\maketitle
	

\section{Introduction}

Two coupled waveguides are essential components of several optical devices such as  optical directional couplers, multicore fibers, polarization splitters, interferometers, and ring resonators \cite{Snyder1983,Marcuse1989, Okamoto2006}.
It is known that the overlap of the modes of evanescently coupled waveguides or cavities results in an optical gradient force  \cite{Povinelli2005,Rakich2009,Roels2010,Rodriguez2015,Rodrigues2017,Miri2018}, and examples of devices where this is exploited include 
coupled strip waveguides \cite{Povinelli2005}, a waveguide suspended over a silica substrate \cite{Rodrigues2017},  coupled slab waveguides \cite{Miri2018}, coupled whispering-gallery-mode microspheres   \cite{Ng2005,Povinelly2005b}, coupled guiding mirrors \cite{Mizrahi2005}, and coupled microring resonators \cite{Rakich2007}.
It has already been shown that the optical gradient force between two coupled dielectric structures is attractive or repulsive depending on whether a symmetric or antisymmetric mode is excited \cite{Povinelli2005,Rakich2009,Roels2010,Rodriguez2015,Rodrigues2017,Miri2018,Ng2005,Povinelly2005b,Mizrahi2005,Rakich2007}.
The optical gradient forces between coupled macroscopic structures have been experimentally observed for a waveguide coupled to a high-$Q$ microresonator \cite{Eichenfield2007} or to a dielectric substrate \cite{Li2008},  coupled nanophotonic waveguides \cite{Li2009,Roels2009}, and coupled ring resonators \cite{Wiederhecker2009}.

Optical devices based on coupled tapered thin fibers have been produced and studied \cite{Birks1995,Glorieux2019,Tong2021}.
Recently, miniaturized optical devices composed of coupled twisted and knotted optical nanofibers have been fabricated \cite{Glorieux2019}. Optical nanofibers have a subwavelength diameter and significantly different core and cladding refractive indices \cite{TongNat03}. Such ultrathin fibers allow for a guided light field, which has tight radial confinement, to propagate along the fiber for a long distance and to interact efficiently with nearby emitters, absorbers, and scatterers \cite{review2016,review2017,Nayak2018}.

Coupling between two optical nanofibers has been studied in the framework of the coupled mode theory \cite{Glorieux2019,CMT}.  A rigorous theory for the guided normal modes of two coupled dielectric rods has been developed using the circular harmonics expansion method  \cite{Wijngaard1973}. This theory 
has been extended to multicore fibers \cite{Yamashita1985,Kishi1989,Huang1990,Chang1997a} and
has been used to study the propagation constant \cite{Wijngaard1973,Chang1997a,Huang1989}, the flux density \cite{Wijngaard1973}, the polarization pattern \cite{Chang1997a}, the mode cutoff   \cite{Chang1997b}, 
the spatial field intensity distributions \cite{tfexact}, and the  atom trapping \cite{tftrap}. 

In this work, we study the optical force between two coupled identical parallel optical nanofibers. We show that the forces  are attractive for even array modes and repulsive for odd array modes. We examine the dependencies of the optical forces on the array mode type, the fiber radius, the light wavelength, and the fiber separation distance. 

The paper is organized as follows. In Sec. \ref{sec:model} we present the model for the system of two coupled identical parallel optical nanofibers and review the calculation methods for the optical forces between the waveguides. In Sec. \ref{sec:num}, we calculate numerically the optical force between the nanofibers. Finally, we conclude in Sec. \ref{sec:summary}.

\section{Two coupled identical parallel optical nanofibers}
\label{sec:model}

\subsection{Model and array modes}

\begin{figure}[tbh]
	\begin{center}
		\includegraphics{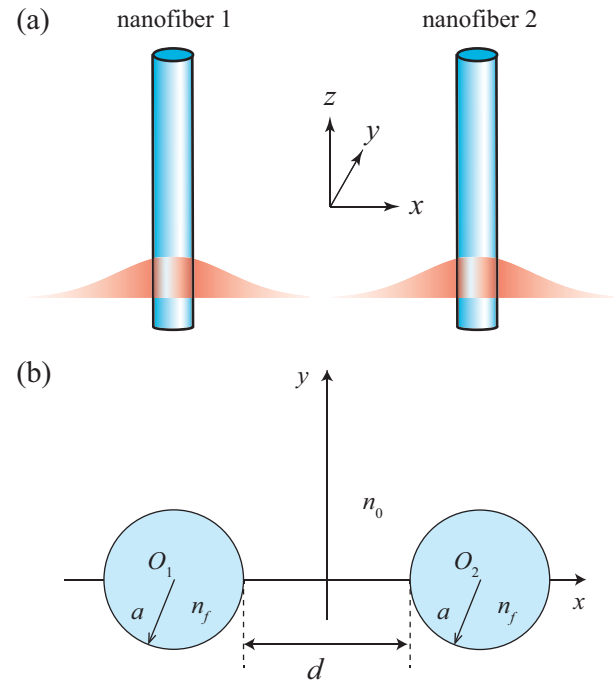}
	\end{center}		
	\caption{(a) Two coupled parallel optical nanofibers and (b) the geometry of the system. 	 
	}
	\label{fig1}
\end{figure}

We study two identical vacuum-clad, optical nanofibers that lie parallel to each other along the direction of the fiber axis  (see  Fig.~\ref{fig1}). The fibers are labeled by the indices $j=1,2$. Each nanofiber $j$ can be viewed as a dielectric cylinder with a radius $a$ and a refractive index $n_f>1$, surrounded by an infinite background of vacuum or air  with a refractive index $n_0=1$. The diameter of each nanofiber is taken to be a few hundreds of nanometers and each 
can support either a single or multiple modes depending on the fiber size parameter $V=ka\sqrt{n_f^2-n_0^2}$. Here, $k=\omega/c$ is the wave number of light with an optical frequency $\omega$ in free space.  

We introduce the global Cartesian coordinate system $\{x,y,z\}$, where $z$ is parallel to the  $z_1$ and $z_2$ axes of the fibers, $x$ is perpendicular to the $z$ axis and connects the centers $O_1$ and $O_2$ of the fibers, and $y$ is perpendicular to the  $x$ and $z$ axes (see Fig.~\ref{fig1}). The plane $xy$ is the transverse (cross-sectional) plane of the fibers. The positions of the fiber centers $O_1$ and $O_2$ along $x$ are taken to be $O_1=-(a+d/2)$ and $O_2=a+d/2$, where $d$ is the separation distance between the two fibers. 
 
The normal modes  of the coupled fibers are termed array modes. We study the array modes of a light field with an optical frequency $\omega$ which propagates in the $+z$ direction with a propagation constant $\beta$. The electric and magnetic components of the field can be written as $\mathbf{E}=[\boldsymbol{\mathcal{E}}e^{-i (\omega t-\beta z)}+\mathrm{c.c.}]/2$ and $\mathbf{H}=[\boldsymbol{\mathcal{H}}e^{-i (\omega t-\beta z)}+\mathrm{c.c.}]/2$, respectively, where $\boldsymbol{\mathcal{E}}$ and $\boldsymbol{\mathcal{H}}$ are the slowly varying complex envelopes. 

The rigorous theory for the guided normal modes of two parallel dielectric cylinders has been formulated in Ref.~\cite{Wijngaard1973}. The results of this theory have been summarized and discussed in detail in Ref.~\cite{tfexact}.
According to Refs.~\cite{Wijngaard1973,tfexact}, two coupled identical parallel fibers
have four types of guided normal modes, namely the even $\mathcal{E}_z$-cosine modes, the odd $\mathcal{E}_z$-cosine modes,
the even $\mathcal{E}_z$-sine modes, and the odd $\mathcal{E}_z$-sine modes.
The cross-sectional profiles of the electric intensity distributions $|\boldsymbol{\mathcal{E}}|^2$ of the fields in these normal modes are illustrated in Fig.~\ref{fig2} and are discussed in Sec. \ref{sec:num}.

\subsection{Optical force between the nanofibers}

A simple dispersion formula for the optical force between coupled parallel waveguides has been derived in Ref.~\cite{Povinelli2005}.
According to Ref.~\cite{Povinelli2005}, an adiabatic change in the separation distance $d$  shifts the eigenmode frequency $\omega$ and results in the optical force per unit length
\begin{equation}\label{f5}
	F=-\bigg(\frac{\partial U}{\partial d}\bigg)_{\beta}=-\frac{U}{\omega}\bigg(\frac{\partial \omega}{\partial d}\bigg)_{\beta}
\end{equation}
between the waveguides.
Here, $U=N\hbar\omega$ is the energy per unit of propagation length of the field in an eigenmode of the combined system, with $N$ being the corresponding number of photons.  
The derivative in Eq.~(\ref{f5}) is taken at a fixed propagation constant $\beta$ to ensure the translational invariance of the optical mode in the longitudinal direction. Negative and positive values of $F$ correspond to attractive and repulsive forces.

Assume that the dispersion relation for the system is $\Phi(\omega,\beta,d)=0$. The triple product (Euler's chain) rule reads $(\partial\omega/\partial\beta)_{d} (\partial\beta/\partial d)_{\omega}
(\partial d/\partial \omega)_{\beta}=-1$.
On the other hand, the optical power transmitted through the system is given as $P=v_gU$, where $v_g=(\partial\omega/\partial\beta)_{d}$ is the group velocity. 
Hence, Eq.~(\ref{f5}) yields the dispersion formula \cite{Povinelli2005}
\begin{equation}\label{f6}
	F=\frac{P}{c}\bigg(\frac{\partial n_{\mathrm{eff}}}{\partial d}\bigg)_{\omega},
\end{equation}
where $n_{\mathrm{eff}}=\beta/k$ is the effective refractive index.

The optical force between the nanofibers can also be calculated from the  Maxwell stress tensor.
The components $T_{ii'}$ of the cycle-averaged Maxwell stress tensor $\mathbf{T}$ are given as \cite{Jackson}
\begin{eqnarray}\label{f7}
	T_{ii'}&=& \frac{1}{4}\mathrm{Re}[
	\epsilon_0n_{\mathrm{ref}}^2(2\mathcal{E}_i \mathcal{E}_{i'}^*-\delta_{ii'}|\boldsymbol{\mathcal{E}}|^2)
		\nonumber\\&&\mbox{}
	+\mu_0(2\mathcal{H}_i \mathcal{H}_{i'}^*-\delta_{ii'}|\boldsymbol{\mathcal{H}}|^2)],
\end{eqnarray}
where $n_{\mathrm{ref}}=n_f$ inside a fiber $j$ and $n_{\mathrm{ref}}=n_0$ outside the fibers.
It is known that the cycle-averaged optical force $\boldsymbol{\mathcal{F}}$ on a body is  \cite{Jackson} 
\begin{equation}\label{f8}
	\boldsymbol{\mathcal{F}}=\oint_{S} \mathbf{T}\cdot d\mathbf{S},
\end{equation}
where $S$ is a closed surface surrounding the body. 
In the case of the coupled parallel nanofibers (see Fig.~\ref{fig1}), only the component $\mathcal{F}_{x}$ of the force $\mathcal{F}$ is nonzero. The optical force per unit of length between the nanofibers is given by $F=\mathcal{F}_{x}/L$,
where $L$ is the length of the nanofibers. We find 
\begin{equation}\label{f9}
	F=-\int\limits_{-\infty}^{\infty} T_{xx} dy,
\end{equation}
where
\begin{eqnarray}\label{p6}
	T_{xx} &=& \frac{1}{4}[
	\epsilon_0n_0^2(2|\mathcal{E}_x|^2-|\boldsymbol{\mathcal{E}}|^2)
	+\mu_0(2|\mathcal{H}_x|^2-|\boldsymbol{\mathcal{H}}|^2)]\qquad
\end{eqnarray}
and the integration is taken along the $y$ axis (see Fig.~\ref{fig1}).
It has been shown for a variety of coupled systems that the results of the calculations for the force from Eq.~(\ref{f6}) are identical to that of the calculations from Eq.~(\ref{f9}) \cite{Povinelli2005,Rodrigues2017,Miri2018}.

\section{Numerical calculations}
\label{sec:num}

\begin{widetext}

\begin{figure}[tbh]
	\begin{center}
		\includegraphics[width=0.9\textwidth]{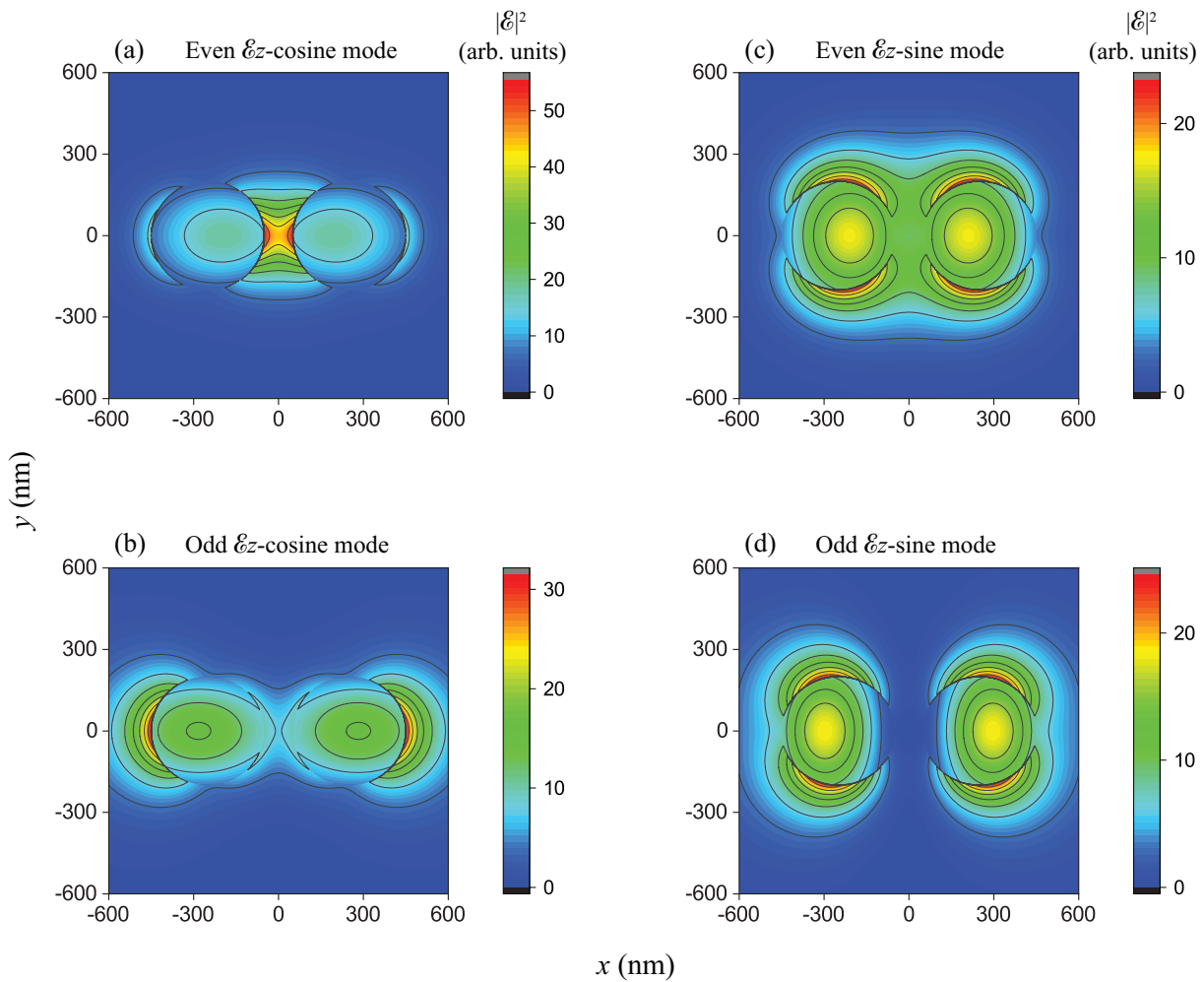}
	\end{center}
	\caption{Cross-sectional profiles of the electric intensity distributions 
		$|\boldsymbol{\mathcal{E}}|^2$ of the fields in (a) the even $\mathcal{E}_z$-cosine mode, (b) the odd $\mathcal{E}_z$-cosine mode, (c) the even $\mathcal{E}_z$-sine mode, and (d) the odd $\mathcal{E}_z$-sine mode 
		of two identical parallel nanofibers. The fiber radius is $a=200$ nm,  the wavelength of light is $\lambda=800$ nm, and the separation distance between the two fibers is $d=100$ nm. The refractive index of the fibers is $n_1=n_2=1.4533$ and that of the surrounding medium is $n_0=1$. The power of light is the same for the modes.}
	\label{fig2}
\end{figure}

\end{widetext}

In this section, we numerically calculate the optical force $F$ produced by the field in a guided array mode of two identical parallel vacuum-clad  silica nanofibers. The refractive index of the vacuum cladding is, as already mentioned, $n_0=1$. To calculate the refractive index $n_1=n_2$ of the silica cores of the nanofibers, the four-term Sellmeier formula for fused silica is used \cite{Malitson,Ghosh}. In particular, for light with the wavelength $\lambda=800$ nm, we have $n_1=n_2=1.4533$.

According to the previous section, in the case of identical fibers, there are four kinds of normal modes, denoted  
as even $\mathcal{E}_z$-cosine, odd $\mathcal{E}_z$-cosine, even $\mathcal{E}_z$-sine, and odd $\mathcal{E}_z$-sine modes \cite{Wijngaard1973}. We are interested in the case where the fiber radius is small enough that no more than one normal mode of each kind can be supported by the fibers.

The spatial distributions of the fields in the guided array modes of two parallel fibers have been studied in Refs.~\cite{Wijngaard1973,tfexact}. In the present paper, we are interested in the optical forces between two nanofibers. These forces can be calculated from the gap-dependent effective refractive index $n_{\mathrm{eff}}$ (or the propagation constant $\beta$) for the array mode using Eq.~(\ref{f6}). They can also be obtained from the field distributions using Eqs.~(\ref{f9}) and (\ref{p6}). 

We use the technique of Refs.~\cite{Wijngaard1973,tfexact} to calculate the field distributions of the guided array modes. 
In Fig.~\ref{fig2}, we plot the cross-sectional profiles of the electric intensity distributions $|\boldsymbol{\mathcal{E}}|^2$ of the fields in the even $\mathcal{E}_z$-cosine, odd $\mathcal{E}_z$-cosine, even $\mathcal{E}_z$-sine, and odd $\mathcal{E}_z$-sine modes. 
We see that $|\boldsymbol{\mathcal{E}}|^2$ is symmetric with respect to the principal axes $x$ and $y$. 
Figure \ref{fig2}(a) shows that the field intensity of the even $\mathcal{E}_z$-cosine mode is dominant in the area between the fibers. 
We observe from Fig.~\ref{fig2}(b) that the field intensity of the odd $\mathcal{E}_z$-cosine mode is dominant in the outer vicinities of the left-side surface of the left-side fiber and the right-side surface of the right-side fiber.
Figure \ref{fig2}(c) shows that the field intensity of the even $\mathcal{E}_z$-sine mode is dominant in the outer vicinities of the top and bottom parts of the surfaces of the fibers, and is significant in the area between the fiber surfaces. 
According to Fig.~\ref{fig2}(d), the field intensity of the odd $\mathcal{E}_z$-sine mode 
is dominant in the vicinities of the top and bottom parts of the surfaces of the fibers, significant in the outer vicinities of the left-side surface of the left-side fiber and the right-side surface of the right-side fiber, and small in the area between the fibers.

It is clear from Fig.~\ref{fig2} that, for each of the two interfacing nanofibers, the field intensity on the inward side (the right side of the left-side nanofiber or the left side of the right-side nanofiber) is different from the field intensity on the outward side (the left side of the left-side nanofiber or the right side of the right-side nanofiber).  
In the case of even array modes, the field  intensity on the inward sides is stronger than that on the outward sides, while in the case of odd array modes, the relation is opposite. It is known that a polarizable microparticle with a dipole induced by a laterally varying optical field will be accelerated towards the region with a stronger field \cite{microparticle}. Nanofibers can be considered as a collection of microscopic dipolar subunits and, hence, can also be accelerated towards the regions with a stronger field. Consequently, we expect that the two nanofibers will be attracted toward or repulsed from each other by the field in an even or odd array mode, respectively.

\begin{figure}[tbh]
	\begin{center}
		\includegraphics{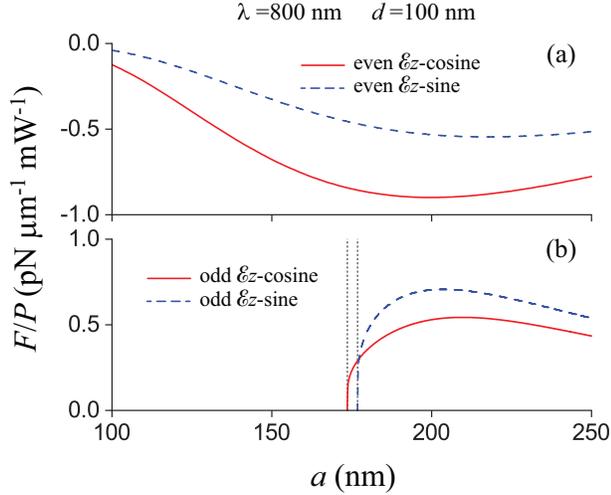}
	\end{center}
	\caption{Power-normalized optical force per unit length $F/P$ as a function of the fiber radius $a$ in the cases of (a) even and (b) odd array modes. The wavelength of light is $\lambda=800$ nm and the separation distance between the two fibers is $d=100$ nm. The refractive index of the fibers is $n_f=1.4533$ and that of the surrounding medium is $n_0=1$. The vertical dotted lines indicate the positions of the cutoffs for the odd array modes. 
	}
	\label{fig3}
\end{figure}

\begin{figure}[tbh]
	\begin{center}
		\includegraphics{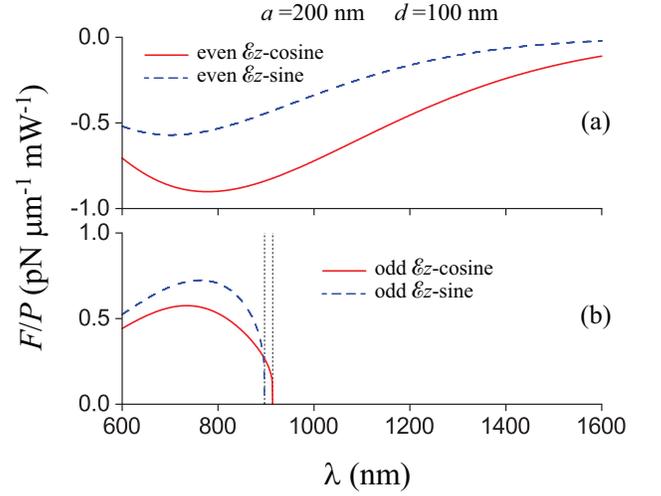}
	\end{center}
	\caption{
		Power-normalized optical force per unit length $F/P$ as a function of the wavelength $\lambda$ of light in the cases of (a) even and (b) odd array modes. The fiber radius is $a=200$ nm and the separation distance between the two fibers is $d=100$ nm. 
		The refractive index $n_f$ of the nanofibers is calculated from the four-term Sellmeier formula for fused silica \cite{Malitson,Ghosh} and that of the surrounding medium is $n_0=1$.  The vertical dotted lines indicate the positions of the cutoffs for the odd array modes.   
	}
	\label{fig4}
\end{figure}

We use the dispersion relation (\ref{f6}) to calculate the optical force $F$ as functions of the fiber radius $a$, the light wavelength $\lambda$, and the separation distance $d$. We note that the use of expression (\ref{f9})
for $F$ in terms of the Maxwell stress tensor gives the same numerical results (see Fig.~\ref{fig8}).

We plot in Figs.~\ref{fig3} and \ref{fig4} the power-normalized optical force per unit length $F/P$ as functions of the fiber radius $a$ and the light wavelength $\lambda$.	
The figures show that the magnitude of the optical force depends on the mode type and the sign of the force depends on the mode parity. Indeed, the forces of the even modes are negative (see the upper parts of the figures), while the forces of the odd modes are positive (see the lower parts of the figures). Thus, the optical forces between the nanofibers are attractive for the fields in the symmetric modes and repulsive for the fields in the antisymmetric modes, in agreement with the results for other systems of coupled  waveguides and cavities \cite{Povinelli2005,Rakich2009,Roels2010,Rodriguez2015,Rodrigues2017,Miri2018,Ng2005,Povinelly2005b,Mizrahi2005,Rakich2007}.  These features are the consequences of the fact that, for increasing fiber separation distance $d$, the propagation constant $\beta$ and, hence, the effective refractive index $n_{\mathrm{eff}}=\beta/k$ decrease in the case of the even  array modes and increase in the case of the odd array modes \cite{tfexact}. The negative and positive signs of the optical forces are also in agreement with the field intensity distributions shown in Fig.~\ref{fig2}, where the nanofibers are expected to be accelerated towards the regions with a stronger field.
Note that the typical order of magnitude of the power-normalized force per unit length $F/P$ for coupled nanofibers is 1 pN $\mu\mathrm{m}^{-1}$ mW$^{-1}$, similar to that for coupled silicon strip waveguides \cite{Povinelli2005}, a silicon waveguide suspended over a silica substrate \cite{Rodrigues2017}, and coupled silicon slab waveguides \cite{Miri2018}.
According to Ref.~\cite{Povinelli2005}, the electrostatic force due to trapped or induced charges in coupled silicon strip waveguides is at least one order smaller than the optical force, and the Casimir-Liftshitz is even smaller. We expect that similar relations between the forces can be realized in the case of nanofibers with an appropriate and reasonable power of light.
We also note that, for the power of 1 mW, the optical force between the nanofibers
is about the same as the electrostatic force between nanomechanical optical fibers with integrated electrodes \cite{Podoliak}.

We observe from Figs.~\ref{fig3} and \ref{fig4} that the odd $\mathcal{E}_z$-cosine and odd $\mathcal{E}_z$-sine modes have cutoffs but the even $\mathcal{E}_z$-cosine and even $\mathcal{E}_z$-sine modes have no cutoff \cite{Wijngaard1973,Chang1997b}. According to Ref.~\cite{tfexact}, the cutoff values of the fiber radius $a$ and the light wavelength $\lambda$ for the odd $\mathcal{E}_z$-cosine and $\mathcal{E}_z$-sine modes depend on the separation distance $d$ between the two fibers. A smaller $d$ leads to a larger cutoff value of the fiber radius $a$ and to a smaller cutoff value of the light wavelength $\lambda$ \cite{tfexact}. 

Figures \ref{fig3} and \ref{fig4} show that the dependencies of the power-normalized force $F/P$  on the fiber radius $a$ and the light wavelength $\lambda$ are not monotonic: the absolute value $|F|/P$ of the power-normalized force has a local maximum as functions of $a$ and $\lambda$. It is clear that, when $a$ is large enough or $\lambda$ is small enough, the magnitude $|F|/P$ of the normalized force reduces with increasing $a$ or decreasing $\lambda$. Similarly,
when $a$ is small enough or $\lambda$ is large enough, the magnitude $|F|/P$ of the normalized force reduces with decreasing $a$ or increasing $\lambda$ in the case of even modes, or has a cutoff in the case of odd modes. 
Comparison between the solid red and dashed blue curves in Figs.~\ref{fig3}(a) and \ref{fig4}(a) shows that the absolute value of the power-normalized force per unit length $F/P$ for the even $\mathcal{E}_z$-cosine mode (solid red  curves) is larger than that for the even $\mathcal{E}_z$-sine mode (dashed blue curves).

\begin{figure}[tbh]
	\begin{center}
		\includegraphics{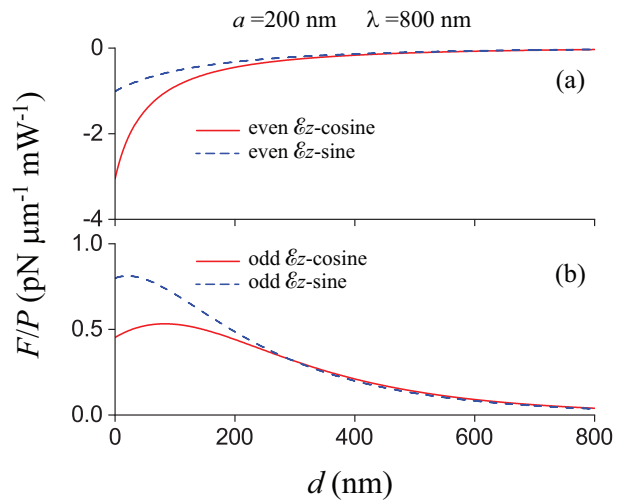}
	\end{center}
	\caption{
		Power-normalized optical force per unit length $F/P$ as a function of the fiber separation distance $d$ in the cases of (a) even and (b) odd array modes. 	
		The fiber radius and the light wavelength are $a=200$ nm and $\lambda=800$ nm.
		The refractive indices of the fibers and the surrounding medium are as in Fig.~\ref{fig3}. 
	}
	\label{fig5}
\end{figure}

\begin{figure}[tbh]
	\begin{center}
		\includegraphics{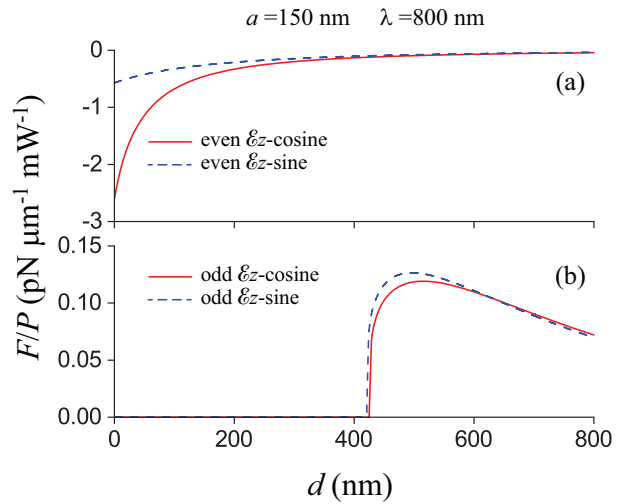}
	\end{center}
	\caption{
		Same as Fig.~\ref{fig5} except for $a=150$ nm.		
	}
	\label{fig6}
\end{figure}

\begin{figure}[tbh]
	\begin{center}
		\includegraphics{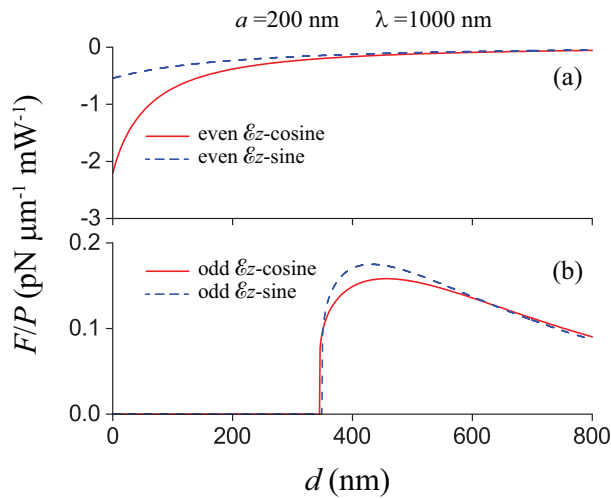}
	\end{center}
	\caption{
		Same as Fig.~\ref{fig5} except for $\lambda=1000$ nm.		
	}
	\label{fig7}
\end{figure}

We plot in Figs.~\ref{fig5}--\ref{fig7} the power-normalized optical force per unit length $F/P$ as functions of the fiber separation distance $d$ for three different sets of values of the fiber radius $a$ and the light wavelength $\lambda$. We observe from Fig.~\ref{fig5} that, when the fiber radius $a$ is large enough or, equivalently, the light wavelength is small enough, there is no cutoff of the guided normal modes for any separation distance $d$. 
However, Figs.~\ref{fig6}(b) and \ref{fig7}(b) show that, if the fiber radius $a$ is small enough or, equivalently, the light wavelength is large enough, a cutoff of an odd guided normal mode may appear at a nonzero fiber separation distance $d$.
We observe from Figs.~\ref{fig5}--\ref{fig7} that the dependence of the normalized force $F/P$  on the fiber separation distance $d$ is monotonic for the even array modes but not monotonic for the odd array modes.

Figures \ref{fig5}--\ref{fig7} show that the differences between the optical forces for the $\mathcal{E}_z$-cosine and $\mathcal{E}_z$-sine modes of the same even or odd parity reduce with increasing fiber separation distance $d$. This feature arises as a consequence of the fact that 
the differences between the optical forces for the $\mathcal{E}_z$-cosine and $\mathcal{E}_z$-sine array modes
are determined by the differences between the propagation constants of the array modes and, hence, by the strength of the coupling between the nanofibers. This coupling depends on the mode overlap and hence reduces with increasing separation distance $d$. In addition, the dependence of the coupling on the mode polarization is very weak in the limit of large $d$ \cite{CMT,tfexact}.

\begin{figure}[tbh]
	\begin{center}
		\includegraphics{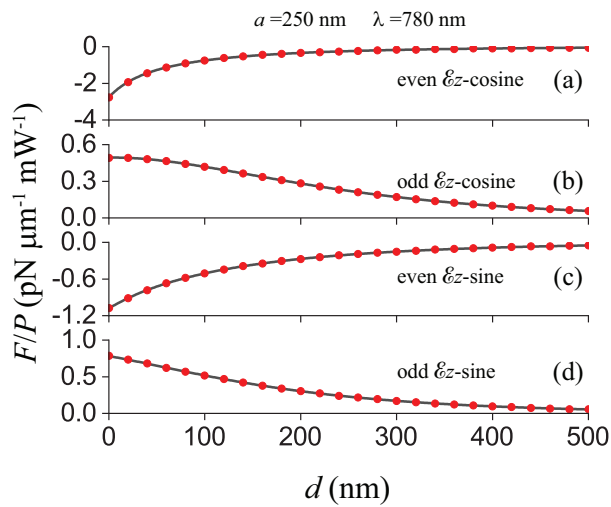}
	\end{center}
	\caption{
		Comparison between the results of the numerical calculations for the
		power-normalized optical force per unit length $F/P$ using Eq.~(\ref{f6}) (black lines) and Eq.~(\ref{f9}) (red points). 	
		The fiber radius and the light wavelength are $a=250$ nm and $\lambda=780$ nm.
		The refractive indices of the fibers and the surrounding medium are as in Fig.~\ref{fig3}. 
	}
	\label{fig8}
\end{figure}

We compare in Fig.~\ref{fig8} the results of the numerical calculations for the
power-normalized optical force per unit length $F/P$ using the dispersion relation (\ref{f6}) (black lines) and the Maxwell stress tensor relation (\ref{f9}) (red points). 	
The figure shows that the two methods give the same numerical results for the optical force  \cite{Povinelli2005,Rodrigues2017,Miri2018}.

\section{Summary}
\label{sec:summary}

We have studied the optical force between two coupled identical parallel  nanofibers using the rigorous array mode theory.
We have shown that the forces of the even array modes are negative (attractive), while the forces of the odd array modes are positive (repulsive). We have investigated the dependencies of the optical forces on the array mode type, the fiber radius, the light wavelength, and the fiber separation distance. We have shown that the dependencies of the optical forces on the fiber radius and the light wavelength are, in general, nonmonotonic. For a given power and a given separation distance, the absolute value of the force achieves a peak when the fiber radius and the light wavelength are appropriate. Our results are important for controlling and manipulating the optical forces between coupled nanofibers.

\begin{acknowledgments}
This work was supported by the Okinawa Institute of Science and Technology (OIST) Graduate University and by the Japan Society for the Promotion of Science (JSPS) Grant-in-Aid for Scientific Research (C) under Grants 19K05316 and 20K03795.
S.N.C. acknowledges support by Investments for the Future from LabEx PALM (ANR-10-LABX-0039-PALM).
The authors gratefully acknowledge J. Twamley for useful discussions.
\end{acknowledgments}


\end{document}